\documentclass[
    ,final            % use final for the camera ready runs
  ]
  {aipproc}

\layoutstyle{6x9}

\begin{document}

\title{Soft and Hard Scale QCD Dynamics in Mesons}


\author{T. Nguyen}{
  address={Center for Nuclear Research, Department of Physics,  Kent State 
University, Kent OH 44242, USA}
}

\author{N. A. Souchlas}{
  address={Center for Nuclear Research, Department of Physics,  Kent State 
University, Kent OH 44242, USA}
}

\author{P. C. Tandy}{
  address={Center for Nuclear Research, Department of Physics,  Kent State 
University, Kent OH 44242, USA}
 %pct% ,altaddress={<author1 address>} % additional visiting address
}

\begin{abstract}
Using a ladder-rainbow kernel previously established for the soft scale of light quark hadrons, we explore the extension to masses and electroweak decay constants of ground state pseudoscalar and vector quarkonia and heavy-light mesons in the c- and b-quark regions.   We make a systematic study of the 
effectiveness of a constituent mass concept as a replacement for a heavy quark dressed propagator.   The difference between vector and axial vector current correlators is examined to estimate the four quark chiral condensate.    The valence quark distributions, in the pion and kaon,  defined in deep inelastic scattering, and measured in the Drell Yan process, are investigated with the same ladder-rainbow truncation of the Dyson-Schwinger and Bethe-Salpeter equations.
\end{abstract}

\maketitle

%%%%%%%%%%%%%%%%%%%%%%%%%%%%%%%%%%%%%%%%%%%%
%% MAINMATTER
%%%%%%%%%%%%%%%%%%%%%%%%%%%%%%%%%%%%%%%%%%%%

\section{DYSON--SCHWINGER EQUATIONS OF QCD}

A great deal of progress in the QCD modeling of hadron physics has been 
achieved through the use of the ladder-rainbow truncation of the Dyson-Schwinger
equations (DSEs).   The DSEs are the equations of motion of a
quantum field theory.  They form an infinite hierarchy of coupled
integral equations for the Green's functions ($n$-point functions) of
the theory.  Bound states (mesons, baryons) appear as poles in the appropriate
Green's functions, and, e.g., the Bethe-Salpeter bound state equation appears after taking residues in the DSE for the appropriate color singlet vertex. 
For recent reviews on the DSEs and their use in hadron physics, see
Refs.~\cite{Roberts:1994dr,Tandy:1997qf,Alkofer:2000wg,Maris:2003vk}.   

In the Euclidean metric that we use throughout, the DSE for the dressed quark propagator is
\begin{eqnarray}
S(p)^{-1}  &=& Z_2 \, i\,/\!\!\!p + Z_4 \, m(\mu) + Z_1 \int^\Lambda_q \! g^2D_{\mu\nu}(p-q) \, 
        \frac{\lambda^i}{2}\gamma_\mu \, S(q) \, \Gamma^i_\nu(q,p)~,
\label{quarkdse}
\end{eqnarray}
where $D_{\mu\nu}(k)$ is the renormalized dressed-gluon propagator,
$\Gamma^i_\nu(q,p)$ is the renormalized dressed quark-gluon vertex.
We use $\int_q^\Lambda$ to denote $\int^\Lambda  d^4q/(2\pi)^4$ with $\Lambda$ being the
mass scale for translationally invariant regularization.    The renormalization condition is 
$S(p)^{-1}=i\gamma\cdot p+m(\mu)$ at a sufficiently large spacelike
$\mu^2$, with $m(\mu)$ the renormalized mass at renormalization scale $\mu$.
We use \mbox{$\mu=19\,{\rm GeV}$}.   The $Z_i(\mu, \Lambda)$ are renormalization constants.  

Bound state pole residues of the inhomogeneous Bethe-Salpeter equation (BSE) for the relevant vertex, yield the homogeneous BSE bound state equation 
\begin{eqnarray}
\Gamma^{a\bar{b}}(p_+,p_-) &=& \int^\Lambda_q \! K(p,q;P)S^a(q_+)
                                          \Gamma^{a\bar{b}}(q_+,q_-)S^b(q_-)~~,
\label{bse}
\end{eqnarray}
where $K$ is the renormalized $q\bar{q}$ scattering kernel that is
irreducible with respect to a pair of $q\bar{q}$ lines.  Quark
momenta are \mbox{$q_+ =$} \mbox{$q+\eta P$} and \mbox{$q_- =$} \mbox{$q -$}
\mbox{$(1-\eta) P$} where the choice of $\eta$ is equivalent to a definition of relative
momentum $q$; observables should not depend on $\eta$.  The meson momentum 
satisfies $P^2 = -M^2$.

%%%%%%%%%%%%%%%%%%%%%%%%%%%%%%%%%%%%%%%%%%%%
%% Sample figure:
%%
%% The option [height=...] scales the picture to the given height,
%% without it it would be printed at its nominal size
%\begin{figure}
% \includegraphics[height=.3\textheight]{golfer}
%  \caption{Picture to fixed height}
%\end{figure}
%%%%%%%%%%%%%%%%%%%%%%%%%%%%%%%%%%%%%%%%%%%%

%\section{\label{sec:RLMT}Rainbow-Ladder Truncation}

A viable truncation of the infinite set of DSEs should respect
relevant (global) symmetries of QCD such as chiral symmetry, Lorentz
invariance, and renormalization group invariance.  For electromagnetic
interactions and Goldstone bosons we also need to respect color singlet  vector 
and axial vector current conservation.  The rainbow-ladder (LR) truncation achieves these ends by
the replacement
\mbox{$K(p,q;P)  \to -4\pi\,\alpha_{\rm eff}(k^2)\, D_{\mu\nu}^{\rm free}(k)
\textstyle{\frac{\lambda^i}{2}}\gamma_\mu \otimes \textstyle{\frac{\lambda^i}{2}}\gamma_\nu $}
along with the replacement of the DSE kernel  by
\mbox{$ Z_1 g^2 D_{\mu \nu}(k) \Gamma^i_\nu(q,p) \to 
 4\pi\,\alpha_{\rm eff}(k^2) \, D_{\mu\nu}^{\rm free}(k)\, \gamma_\nu
                                        \textstyle\frac{\lambda^i}{2} $}
where $k=p-q$, and $\alpha_{\rm eff}(k^2)$ is an effective running
coupling.   This truncation is the first term in a systematic
expansion~\cite{Bender:1996bb,Bhagwat:2004hn} of $K$; asymptotically, it reduces to leading-order perturbation theory.
These two truncations are mutually consistent: together they produce color singlet vector and axial-vector vertices satisfying their respective Ward identities.  This
ensures that the chiral limit ground state pseudoscalar bound states
are the massless Goldstone bosons from chiral symmetry
breaking~\cite{Maris:1998hd,Maris:1997tm}; and
ensures electromagnetic current conservation~\cite{Roberts:1996hh}.  
%%%%%%%%%%%%%%%%%%%%%%%%%%%%%%%%%%%%%%%%%%%%
%% SAMPLE TABLE
%%
%% Shows the use of \tablehead and \tablenote
%% macros

%\begin{table}
%\begin{tabular}{lrrrr}
%\hline
%  & \tablehead{1}{r}{b}{Single\\outlet}
%  & \tablehead{1}{r}{b}{Small\tablenote{2-9 retail outlets}\\multiple}
%  & \tablehead{1}{r}{b}{Large\\multiple}
%  & \tablehead{1}{r}{b}{Total}   \\
%\hline
%1982 & 98 & 129 & 620    & 847\\
%1987 & 138 & 176 & 1000  & 1314\\
%1991 & 173 & 248 & 1230  & 1651\\
%1998\tablenote{predicted} & 200 & 300 & 1500  & 2000\\
%\hline
%\end{tabular}
%\caption{Average turnover per shop: by type
%  of retail organisation}
%\label{tab:a}
%\end{table}
%%%%%%%%%%%%%%%%%%%%%%%%%%%%%%%%%%%%%%%%%%%%
%----------------------------------------------------------------------------------------------------------------------
\begin{table}
\caption{DSE results~\protect\cite{Maris:1999nt} for pseudoscalar and vector meson masses and electroweak decay constants, together with experimental data~\protect\cite{PDG04}.   Units are GeV except where indicated.   Quantities marked by $\dagger$ are fitted with the indicated current quark masses and the infrared strength parameter of the ladder-rainbow kernel.  \label{Table:model} }

\begin{tabular}{|l|cc|cc|cc|cc|cc|} \hline 
  \multicolumn{2}{|c|}{ }   & \multicolumn{3}{c|}{$m^{u=d}_{\mu=1 {\rm GeV}}$}  &  \multicolumn{3}{c|}{$m^{s}_{\mu=1 {\rm GeV}}$}  &   \multicolumn{3}{c|}{- $\langle \bar q q \rangle^0_{\mu=1 {\rm GeV}}$}    \\   \hline
 \multicolumn{2}{|c|}{ expt }   &  \multicolumn{3}{c|}{ 3 - 6 MeV}   &  \multicolumn{3}{c|}{  80 - 130 MeV } &  \multicolumn{3}{c|}{ (0.24 GeV)$^3$ }       \\     
\multicolumn{2}{|c|}{  calc }  &   \multicolumn{3}{c|}{ 5.5 MeV}  &  \multicolumn{3}{c|}{  125 MeV }  &   \multicolumn{3}{c|}{  (0.241 GeV)$^{3\dagger}$  }       \\ \hline
        & $m_\pi$ & $f_\pi$ & $m_K$ & $f_K$   &   $m_\rho$ &  $f_\rho$  & $m_K^\star$ & $f_K^\star$ &  $m_\phi$ &  $f_\phi$
 \\ \hline 
 expt  &   0.138  &  0.131  &   0.496   &   0.160 &   0.770  &  0.216  &   0.892 &  0.225   &   1.020   &  0.236    \\  
 calc  &   0.138$^\dagger$ &  0.131$^\dagger$ & 0.497$^\dagger$ &  0.155  &  0.742 & 0.207 & 0.936 & 0.241  &  1.072     &    0.259   \\ \hline  
\end{tabular}
\end{table}
%----------------------------------------------------------------------------------------------------------------------

We employ the ladder-rainbow kernel found to be successful in earlier 
work for light quarks~\cite{Maris:1997tm,Maris:1999nt}.   It can be written
\mbox{$\alpha_{\rm eff}(k^2) =  \alpha^{\rm IR}(k^2) + \alpha^{\rm UV}(k^2) $}.
The IR term implements the strong infrared enhancement in the region
\mbox{$0 < k^2 < 1\,{\rm GeV}^2$} required for sufficient dynamical
chiral symmetry breaking.   The UV term preserves the one-loop renormalization group behavior of QCD: \mbox{$\alpha_{\rm eff}(k^2) \to \alpha_s(k^2)^{\rm 1 loop}(k^2)$} in the ultraviolet with 
\mbox{$N_f=4$} and \mbox{$\Lambda_{\rm QCD} = 0.234\,{\rm GeV}$}.   The strength of $\alpha^{\rm IR}$ along with two quark masses are fitted to $\langle\bar q q\rangle $, $m_{\pi/K}$.   Selected light quark meson results are displayed in Table~\ref{Table:model}.   The infrared kernel component is phenomenological because QCD is unsolved in such a non-perturbative domain. To help replace such phenomenology by specific mechanisms, it is necessary to first  characterize its performance in new domains.

\section{Heavy Quark Mesons}

In Table~\ref{Table:qQ} we display the results for the heavy-light  ground state pseudoscalars and vectors involving a c-quark or b-quark.   We use DSE solutions for the dressed light quarks.  If a constituent mass propagator is used for the heavy quark, with the constituent mass obtained from a fit to the lightest pseudoscalar, the various meson masses are easily reproduced.   The constituent masses found this way are $M_c^{\rm cons} = 2.0 $~GeV for the c-quark, and $M_b^{\rm cons} = 5.3 $~GeV for the b-quark.   To compare with what is known, we take the quark current 
masses~\cite{PDG04} \mbox{$m_c = 1.2\pm0.2$}~GeV, and \mbox{$m_b = 4.2\pm0.2$}~GeV at scale $\mu =$ 2~GeV and use the quark DSE to run the masses into the timelike 
region where the meson mass shells are located.   If all the meson momentum runs through the heavy quark, the mass function from the DSE would suggest an effective quark mass  
$M_q^{\rm DSE}(p^2\sim -M^2)$ where M is the meson mass.  These DSE masses reproduce the previously obtained values for $M_{c/b}^{\rm cons}$ within 10\%.  In this sense, heavy quark dressing is well summarized by a constituent mass.
However, the electroweak decay constants obtained from the constituent mass approximation are 
30-50\% below the experimental values.    Moreover, within this LR model, quark dressing is not a minor effect because the use of fully dressed quark propagators, both heavy and light, does not yield a physical bound state solution for these heavy-light states involving a c-quark or b-quark.   
%----------------------------------------------------------------------------------------------------------------------
\begin{table}
\caption{Calculated masses and electroweak decay constants for ground state pseudoscalar 
and vector heavy-light mesons, together with experimental data~\protect\cite{PDG04}, all in GeV.   In the rows labelled {\it calc M}, the heavy quark is described by a constituent mass fit to the lightest pseudoscalar (marked by $\dagger$).    In the rows labelled $k_{\rm min}$, the heavy quark is dressed through the DSE with an infrared suppression of the gluon momentum as described in the text.   No such infrared suppression was applied to the dressing of the light quark and the binding kernel.   The bracketed values for $B_s$ and $B_c$ indicate that the entire $\alpha^{\rm IR}(k^2)$ term of the kernel was eliminated, as in earlier work, and this corresponds to a $k_{\rm min}$ value that is about 20\% larger than the value used for $B$. 
\label{Table:qQ} }

\begin{tabular}{|l|cc|cc|cc|cc|cc|} \hline 
      & D & D$^*$ & D$_s$ & D$^*_s$   &   B &  B$^{*}$ & B$_s$ & B$^{*}_s$ &  B$_c$ &  B$^{*}_c$
 \\ \hline 
 expt M &   1.86  &  2.01        &   1.97    &   2.11 &   5.28  &  5.33  &   5.37 &  5.41   &   6.29   &     ?    \\  
 calc M &   1.85$^\dagger$ &  2.04 & 1.97 &  2.17  &  5.27$^\dagger$ & 5.32 & 5.38 & 5.42  &   6.36     &    6.44   \\ 
with $k_{\rm min}$ &  1.88 &    & 1.90  &    &   5.15  &     & (4.75)   &    &  (5.83)  &    \\ \hline  
 expt f   &  0.222                  &      ?     & 0.294 &      ?   &  0.176                  &       ?     &    ?    &    ?    &      ?       &        ?    \\ 
 calc f    & 0.154                  &  0.160 &0.197 & 0.180       & 0.105                  &  0.182 & 0.144 & 0.20 &  0.210   &  0.18      \\ 
 with $k_{\rm min}$ &  0.260 &    & 0.275  &     &   0.265  &     &  (0.164)   &    &  (0.453)  &    \\ \hline  
            
\end{tabular}
\end{table}
%----------------------------------------------------------------------------------------------------------------------

The results for equal quark mesons (quarkonia), displayed in Table~\ref{Table:quarkonia}, show a different perspective.    With the same fitted constituent masses, the quarkonia masses are well reproduced, and again the electroweak decay constants $f_M$ are  too low by some 30-70 \%.    Use of dynamically dressed propagators removes almost all of this deficiency and the decay constants are well reproduced. This improvement provided by dynamical dressing of c- and b-quarks is persistent and systematic.  When the dressing is  progressively  introduced into all three stages (bound state solution, normalization loop integral, and then the loop integral for $f_M$), the result always increases towards the experimental value. 

Our results for decay constants $f_M$ of both heavy quarkonia and heavy-light mesons indicate that a constituent mass approximation, even for b-quarks, is surprizingly inadequate.  This is strongly associated with the very weak binding of such  mesons which magnifies the effect of a weak momentum dependence in the heavy quark mass function $M_q^{\rm DSE}(p^2)$ and the quark field renormalization function $Z_q^{\rm DSE}(p^2)$  in the time-like domain where the mass shell is to be formed.  A mass shell will not form unless all the momentum dependence conspires to push the eigenvalue of the BSE kernel to cross unity. 

%---------------------------------------------------------------------------------------------------------------------
\begin{table}
\caption{Calculated masses and electroweak decay constants for ground state pseudoscalar 
and vector quarkonia, together with experimental data~\protect\cite{PDG04}, all in GeV.  Results employ either a constituent mass treatment of the heavy quark, labelled {\it $ M_Q^{\rm cons}$}, or a dynamical dressed propagator, labelled {\it $ \Sigma_Q^{\rm  DSE}(p^2)$}.    The last  row
shows the results from an effective smooth infrared cut-off of the gluon momentum applied to both binding from the BSE kernel and to quark propagator dressing within the DSE.
\label{Table:quarkonia} }

\begin{tabular}{|l|cc|cc|cc|cc|} \hline 
         & $M_{\eta_c}$  &  $f_{\eta_c}$   & $M_{J/\psi}$  &   $f_{J/\psi}$    & $M_{\eta_b}$  &  $f_{\eta_b}$   & $M_\Upsilon$  &   $f_\Upsilon$     \\ \hline 
 expt                    &   2.98                &     0.340             &      3.09           &  0.411        &   9.4 ?                &            ?           &      9.46           &  0.708              \\ 
 calc  with $ M_Q^{\rm cons}$ &3.02&    0.239          &      3.19           &  0.198    &9.6&    0.244          &      9.65           &  0.210              \\ 
 calc  with $ \Sigma_Q^{\rm  DSE}(p^2)$&3.04&    0.387   &     3.24           &   0.415   &9.59&  0.692   &     9.66           &   0.682    \\   \hline
with  $k_{\rm mim} $  &   3.04  &  0.371  & 3.24  &   0.398    &  9.59  &  0.630  &  9.66   &  0.621 \\   
\hline  
\end{tabular}
\end{table}
%---------------------------------------------------------------------------------------------------------------------
With increasing quark mass, the quarkonia states become smaller in size and the ultraviolet sector of the  ladder-rainbow kernel becomes more influential.   The size of the dressed quark quasi-particle also decreases with quark mass so that the entire heavy quarkonia dynamics becomes increasingly dominated by the ultraviolet  sector of the kernel as dictated by pQCD.  
Although a ladder-rainbow model of $K_{\rm BSE}$ has no dependence on quark or meson masses apart from quark propagators, our results for heavy quarkonia are quite good; one can confirm that the dynamically generated size of the meson wavefunction naturally concentrates the physics increasingly within the ultraviolet sector of the kernel. 

For heavy-light mesons, with increasing heavy quark mass,  the infrared sector of the dynamics experienced by the light quark quasiparticle can be expected to remain dictated by the same soft QCD physics that determined our model kernel.    The question then arises as to whether the self interaction of heavy quark  to produce a quasiparticle, and its coupling via gluon exchange with the light partner, are adequately described by such a kernel.   In a schematic analysis of $K_{\rm BSE}$ beyond ladder-rainbow 
truncation~\cite{Bhagwat:2004hn,Bhagwat:2004kj,Matevosyan:2006bk} it has been shown that an important  contributor to its strength of attraction is dressing of the quark-gluon vertex.  
This mechanism will decrease with increasing quark mass.   However our ladder-rainbow effective kernel was established for light quarks and does not contain such a decreasing strength.
%--------------------------------------------------------------------------------------------
\begin{figure}[ht]
\includegraphics[height=0.26\textheight]{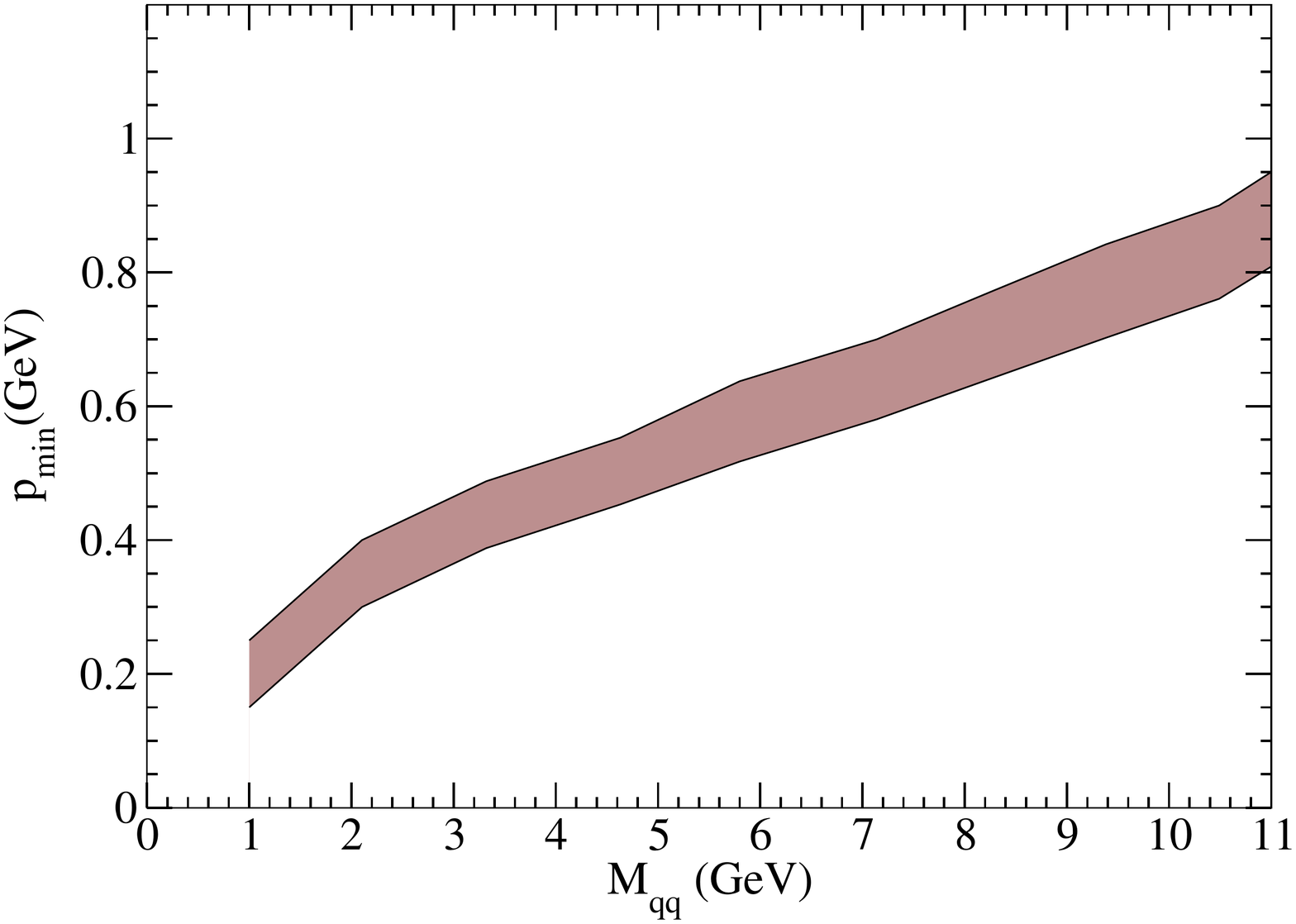}\hspace*{6mm}
\includegraphics[height=0.26\textheight]{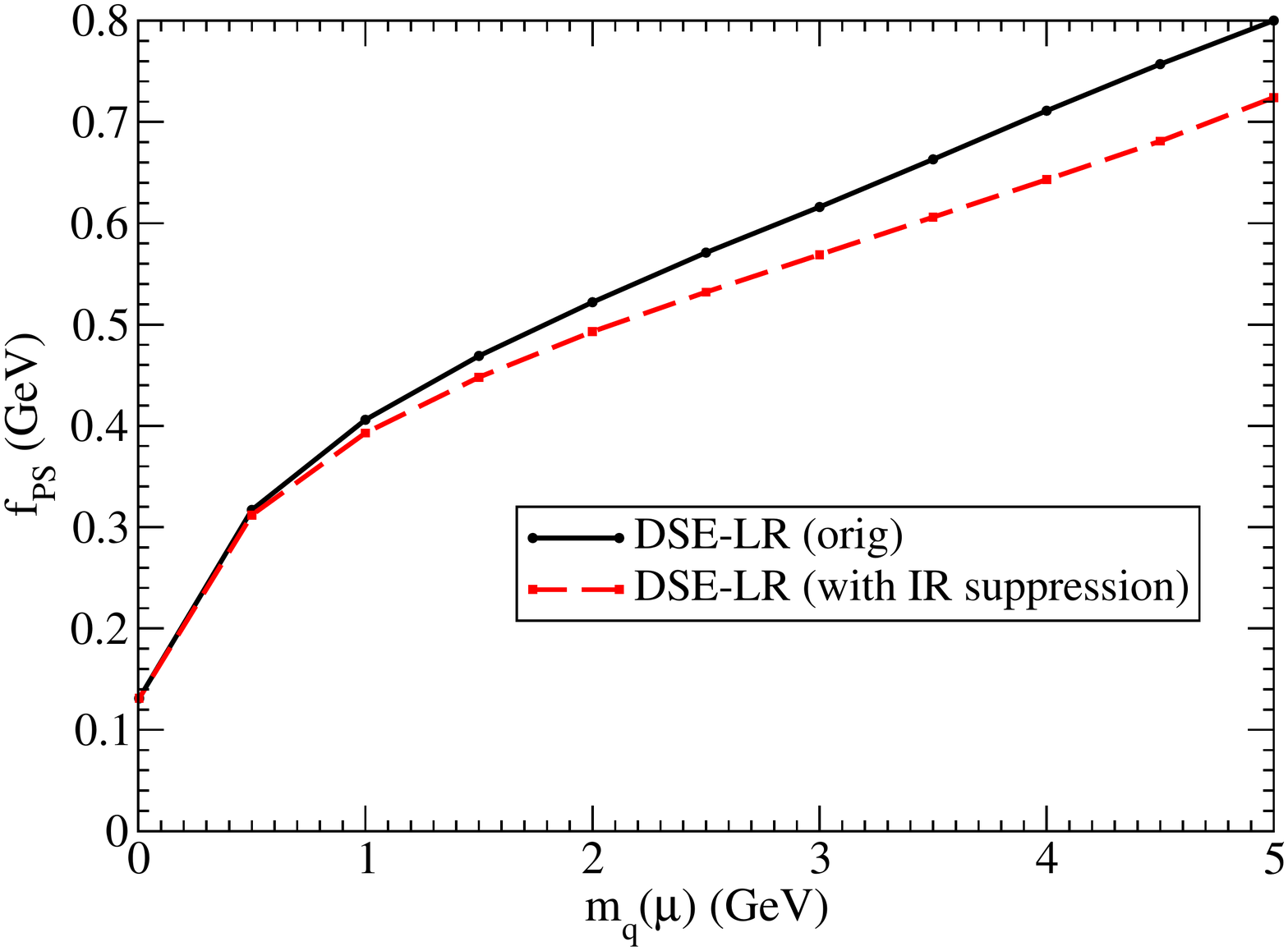}

\caption{{\it Left Panel}:  The effective infrared gluon momentum cutoff versus pseudoscalar quarkonia mass.     {\it Right Panel}:  The pseudoscalar quarkonia electroweak decay constant versus quark current mass.    Solid curve is the is from the original LR kernel without IR suppression; dashed curve is with $m_q$-dependent IR suppression as described in the text. \label{Fig:k_min+psdecay} }
\end{figure}
%--------------------------------------------------------------------------------------------

For this reason we investigated a schematic suppression of the infrared sector of our kernel with increasing quark mass \mbox{$m_q \leq m_b $}, using the form \mbox{$\tilde{\alpha}^{\rm IR}(k^2) = $} \mbox{$f(m_q(\mu))\, \alpha^{\rm IR}(k^2) $} where \mbox{$ f(m_q(\mu)) = $}  \mbox{$ 1 - 0.624 \, m_q(\mu)/ m_b(\mu) $}, the renormalization scale of the model is \mbox{$\mu = 19~{\rm GeV}$}, and \mbox{$m_b(\mu) = 3.8~{\rm GeV} $}.   This smooth infrared suppression can be recast as an effective sharp lower cutoff $k_{\rm min}$ for the gluon momentum and we estimate such a cutoff by equating integrals of the kernel:   \mbox{$\int_k^\Lambda  \tilde{\alpha}^{\rm IR}(k^2) =$}  \mbox{$ \int_{k > k_{\rm min}}^\Lambda \alpha^{\rm IR}(k^2) $}.   The value of $k_{\rm min}$ will be discussed below. This $\tilde{\alpha}^{\rm IR}$ was used for both binding and quark dressing for the quarkonia states, and the results are shown in the indicated row of Table~\ref{Table:quarkonia}.    Only the decay constants show a response and they decrease by less than 10\%, as shown in 
Fig.~\ref{Fig:k_min+psdecay}.

This schematic infrared suppression is not well defined for a meson with two distinct quark mass scales.   So for initial investigation we employed the suppressed kernel for dressing the heavy quark and the original kernel for the dressing of the light quark and the binding to form the meson.
The results displayed in  Table~\ref{Table:qQ} show that physical $D, D_s$ and $B$ states are produced this way, whereas that was impossible with a b-quark dressed via the full kernel.   We have not investigated the vector states this way.   The pseudoscalar masses are quite reasonable; the $D$ and $D_s$ decay constants are closer to experiment than the values obtained with the constituent mass approximation.   We have not tried to fine tune this suppression phenomenology; the $f_B$ value could be improved.    The bracketed results for $B_s$ and $B_c$ were produced by removal of the entire infrared component of our kernel, not by necessity, but for temporary convenience.
   
The results indicate that our ladder-rainbow kernel, determined by essentially chiral quark physics, has implicit infrared strength that should be systematically reduced as quark mass is raised into the b-quark region.   One may think of this as implicit quark-gluon vertex dressing in QCD that ebbs away, as it must, with increasing quark mass.   Another viewpoint is provided by the analysis of  Brodsky and Shrock~\cite{Brodsky:2008be} that, because QCD color confinement limits quarks and gluons to be found only within hadrons, their dynamics is limited by a universal maximum wavelength characteristic of hadron size.   Thus there is a minimum momentum for such virtual fields.    Since the size of hadrons is a dynamically generated scale, exact QCD dynamics could be expected to automatically generate a suppression of modes below a minimum momentum.  However with truncations and models of QCD, such as the one we deal with, an infrared momentum cutoff may need to be a model element to represent explicitly excluded higher order processes.      The  minimum momentum must increase with mass of the hadron or dressed quark/gluon quasiparticle.
A Compton wavelength translates to a rest mass, and thus we expect a minimum momentum to be proportional to mass.   In Fig.~\ref{Fig:k_min+psdecay}, we display the effective sharp $k_{\rm min}$ as a function of  quarkonia mass that corresponds to our smooth infrared suppression investigation.   The shaded band estimates the associated uncertainties.   The increase in proportion to the mass is as expected.    

\section{Leading npQCD scale and the four quark condensate}

Quark helicity and chirality in QCD are increasingly good quantum numbers at short distances or at momentum scales significantly larger than any mass scale.   One manifestation of this is that, for chiral quarks, the correlator of a pair of vector currents is identical to the corresponding correlator of a pair of axial vector currents to all finite orders of pQCD.   Non-perturbatively,  the difference of such correlators measures chirality flips, and the leading non-zero ultraviolet contribution identifies the leading non-perturbative phenomenon in QCD.     This is the four quark 
condensate~\cite{Narison:1989aq}.    The LR kernel can produce the difference correlator as a vacuum polarization integral in momentum space, where the propagators are dressed and the vector and axial vector vertices are generated in a way consistent with the symmetries.   We use the large spacelike momentum dependence to extract the leading coefficient or condensate. 

The vector current-current correlator is formulated as the loop integral
\begin{equation}
\Pi_{\mu\nu}^{V}(P) = \int d^{4}x \; {\rm e}^{iP \cdot x} 
\langle 0|T \, j_\mu(x)\, j^+_\nu(0)|0 \rangle = -\int^\Lambda_q 
Tr\lbrace \gamma_\mu S(q_+) \Gamma^{V}_{\nu} (q,P)S(q_-)\rbrace ~~,
\label{Eq:vcorr}
\end{equation}
where $\Lambda $ indicates regularization, e.g., by the Pauli-Villars method, and $\Gamma^{V}_{\nu}$ is the dressed vector vertex.   The axial vector correlator is formulated in an analogous way and we directly calculate the difference correlator which does not require ultraviolet regularization.   
With \mbox{$\Pi_{\mu\nu}^{V}(P) = (P^{2}\delta_{\mu\nu} - P_{\mu} P_{\nu})\Pi_{T}^{V}(P^{2})$}, and 
\mbox{$\Pi_{\mu\nu}^{A}(P) = (P^{2} \delta_{\mu\nu} - P_{\mu} P_{\nu})
\Pi_{T}^{A}(P^{2})+ P_{\mu} P_{\nu}\, \Pi_{L}^{A}(P^{2}) $}, the quantity of interest here is 
\mbox{$\Pi_{T}^{V-A}(P^{2}) = \Pi_{T}^{V}(P^{2}) - \Pi_{T}^{A}(P^{2}) $}.  

The leading non-perturbative contribution  to $\Pi_{T}^{V-A}$ starts with 
dimension $d =6$ and involves the four-quark condensate in the form~\cite{Dominguez:1998wy,Dominguez:2003dr}
\begin{equation}
\label{eq:c4cm}
\Pi_{T}^{V-A}(P^{2}) = -\frac{32 \pi}{9} 
\frac{\alpha_{s} \langle \bar{q}q\bar{q}q\rangle }{P^{6}}
\{1+\frac{\alpha_{s}(P^{2})}{4\pi}[\frac{247}{12}+ {\rm ln} (\frac{\mu^{2}}{P^{2}})]
\} + O(\frac{1}{P^{8}})~~.
\end{equation}
Our numerical calculation of $P^6\, \Pi_{T}^{V-A}(P^{2})$ identifies a leading ultraviolet constant reasonably well.   The four quark condensate 
$\langle \bar{q}q\bar{q}q\rangle $ extracted via Eq.~(\ref{eq:c4cm}) is 65\% greater than the common vacuum saturation assumption $\langle \bar{q}q\rangle^2$ at the renormalization scale \mbox{$\mu = 19$}~GeV used in this work.  
The low $P^2$ limit provides a reasonable account of the first Weinberg sum rule~\cite{Weinberg:1967kj, Dorokhov:2003kf}:   
\mbox{$P^{2} \,\Pi_{T}^{V-A}(P^{2})|_{P^{2} \to 0}  = - f_{\pi}^{2} $}, in the \mbox{$f_\pi = 0.0924$} GeV convention.    This limit is due to $\Pi_{T}^{A}$ only and  we obtain  \mbox{$f_\pi = 0.09$} GeV that way.   Our results are consistent with the  second Weinberg sum rule~\cite{Weinberg:1967kj} \mbox{$P^4 \,\Pi_{T}^{V-A}(P^{2})|_{P^{2} \to \infty}  =0$}.    The Das-Guralnik-Mathur-Low-Young sum rule~\cite{Das:1967it} relates 
\mbox{$\int_{0}^{\infty}{dP^{2}}\, P^{2}\,\Pi_{T}^{V-A}(P^{2})$} to  the strong component of $m_{\pi^{\pm}} - m_{\pi^{0}}$.   We obtain 4.86 MeV for this mass difference in comparison with $4.43 \pm 0.03$  from experiment.   The chirality-flip ratio 
\mbox{$\Pi_{T}^{V-A}(P^2) / \Pi_{T}^{V+A}(P^2)$}  identifies a scale of $\sim 0.5$~fm for the onset of non-perturbative dynamics~\cite{Nguyen:2009if}.   This DSE model in LR format  produces the same scale as obtained  in both a lattice-QCD calculation and the Instanton Liquid Model~\cite{Faccioli:2003qz}.   

\section{Valence quark distributions in the pion and kaon}

Data for the momentum-fraction probability distributions of quarks and gluons in the pion have primarily been inferred from
Drell-Yan~\cite{Badier:1983mj,Betev:1985pg,Conway:1989fs} and direct photon production~\cite{Aurenche:1989sx} in pion-nucleon and pion-nucleus collisions, and semi-inclusive e$\,$p $\to$ e$\,NX$ reactions~\cite{Adloff:1998yg}.     For a recent review of nucleon and pion parton distributions 
see Ref.~\cite{Holt:2010vj}.
%---------------------------------------------------------------
% figure from Mazatlan talk plus prelim ratio calc after Pade fit
\begin{figure}[th] 
\hspace*{4mm}
\includegraphics[height=0.26\textheight]{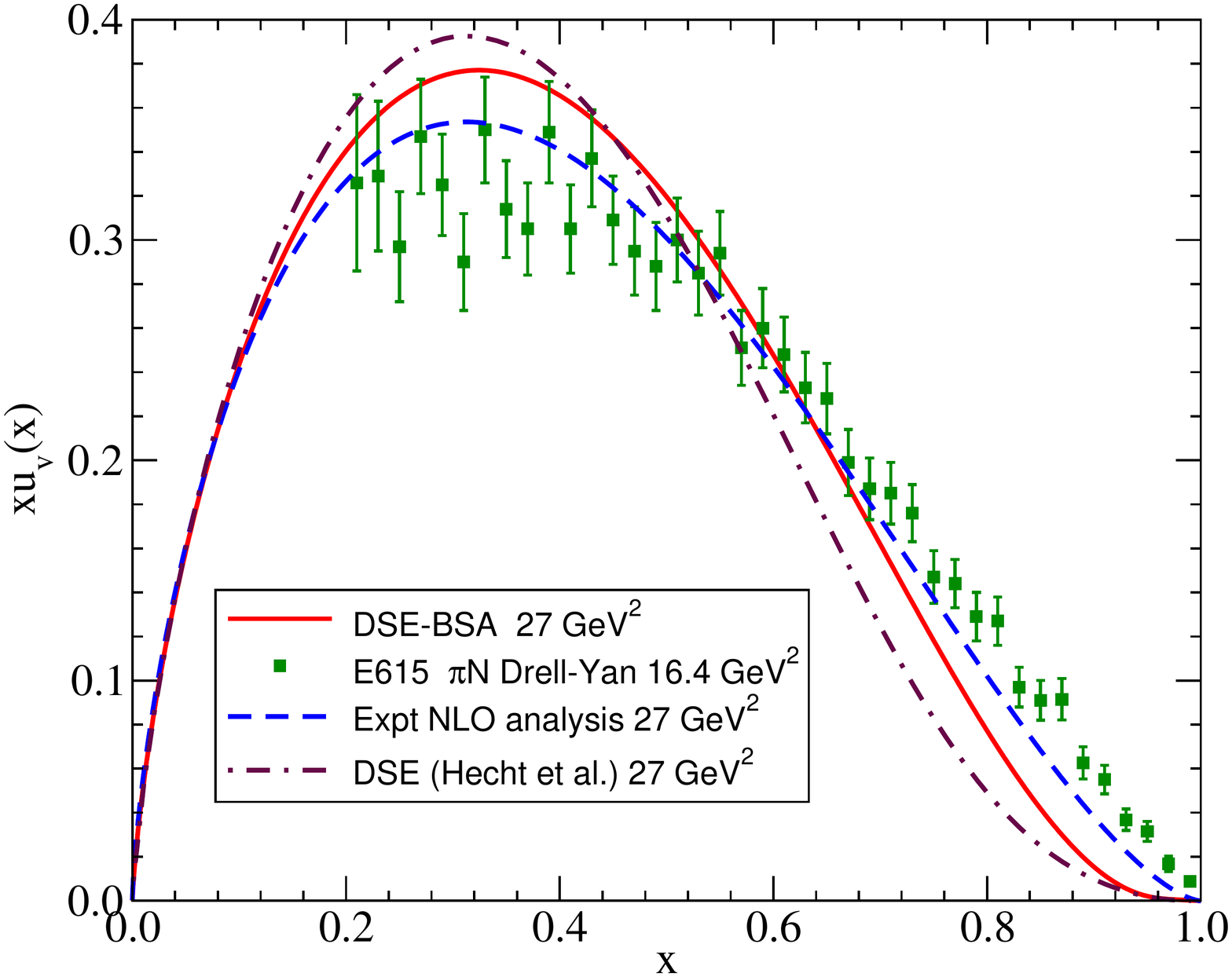}\hspace*{2mm}
\includegraphics[height=0.28\textheight]{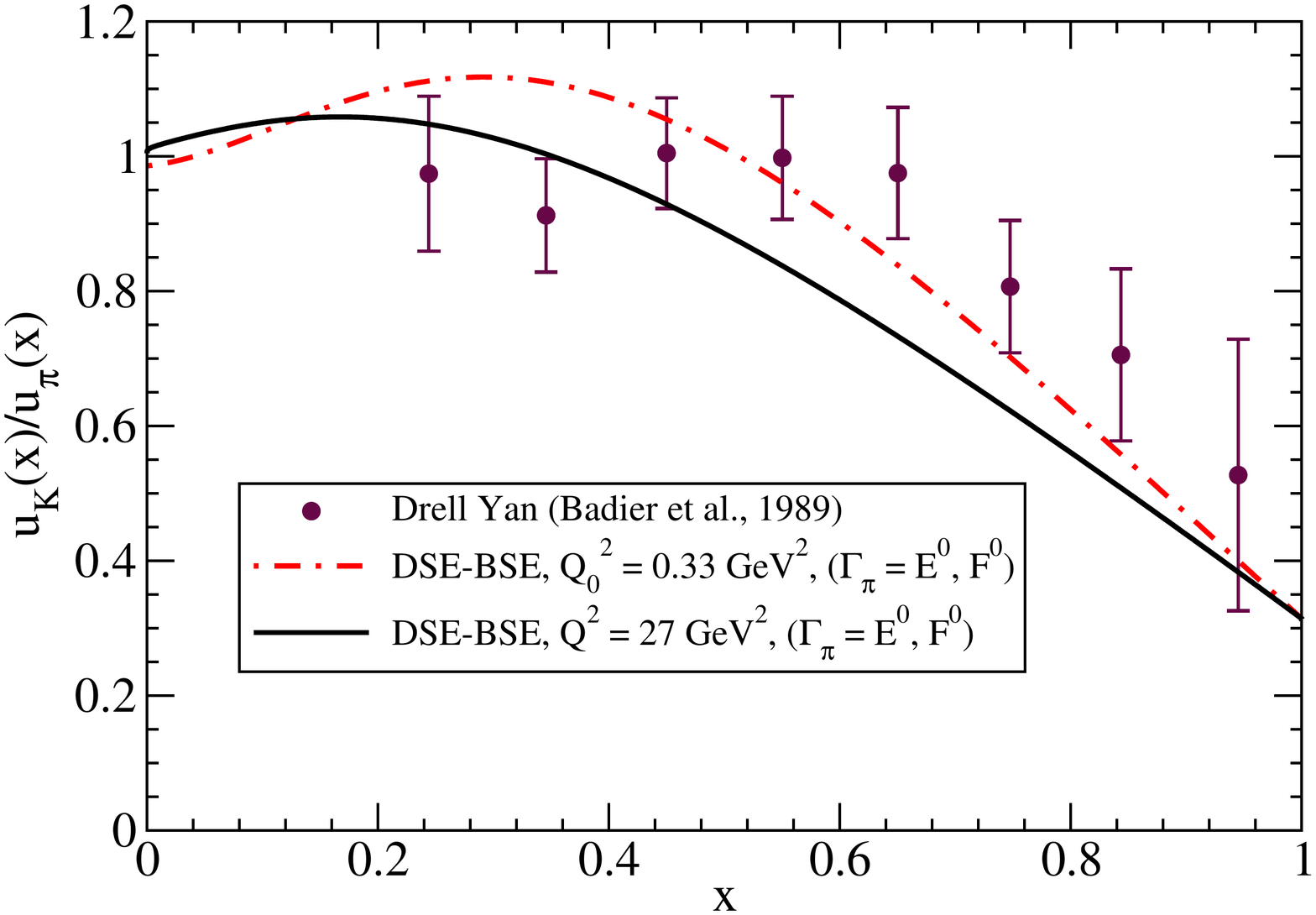}

\caption{{\it Left Panel}:    Pion valence quark distribution evolved to (5.2~GeV)$^2$.  Solid line is the full DSE-BSA calculation~\protect\cite{Nguyen_PhD09}; dot-dashed line is the semi-phenomenological DSE-based calculation of Hecht et al.~\protect\cite{Hecht:2000xa};   experimental data points are from~\protect\cite{Conway:1989fs} at scale (4.05~GeV)$^2$; the dashed line is the recent NLO re-analysis of the experimental data~\protect\cite{Wijesooriya:2005ir}.   {\it Right Panel}:   The ratio of u-quark distributions in the kaon and pion.   The solid line is our preliminary result from DSE-BSE 
calculations~\protect\cite{Nguyen:inprep10,Nguyen_PhD09,Holt:2010vj};  the experimental data is from~\protect\cite{Badier:1980jq,Badier:1983mj}.  \label{fig:pi_DSE+ratio} }
\end{figure} 
%---------------------------------------------------------------

Lattice-QCD is restricted to  low moments of the distributions, not the
distributions themselves~\cite{Best:1997qp}.    Model calculations of deep inelastic scattering (DIS) parton distribution functions are challenging because it is necessary to have perturbative QCD features (including the evolution of scale)  coexisting with a covariant nonperturbative model made necessary by the bound state nature of the target.     Aspects of chiral symmetry have led to DIS calculations within the Nambu--Jona-Lasinio model~\cite{Shigetani:1993dx,Weigel:1999pc,Bentz:1999gx} but there are a number of difficulties with this approach~\cite{Weigel:1999pc,Bentz:1999gx}, among them a point structure for the pion BS amplitude at low model scale and a marked sensitivity to the regularization procedure due to the lack of renormalizability.    Constituent quark models have also been employed~\cite{Szczepaniak:1993uq,Frederico:1994dx}, with the difficulties encountered in such studies considered in Ref.~\cite{Shakin:1994rk}.   Instanton-liquid models~\cite{Dorokhov:2000gu} have also been used.    In these approaches, it is difficult to have pQCD elements join smoothly with nonperturbative aspects.
All of these issues can in principle be addressed if parton distribution functions can be obtained from a model based on the DSEs, and in particular styled after  the approach we have previously made for the related pion electromagnetic form factor~\cite{Maris:2000sk,Maris:1999bh}.    

In the Bjorken kinematic limit, DIS selects the most singular behavior of a correlator of quark fields of the target with light-like distance separation $z^2 \sim 0 $.   With incident photon momentum along the negative 
3-axis, the kinematics selects \mbox{$z^+ \sim z_\perp \sim 0$}  leaving  $z^-$ as the finite distance conjugate to quark momentum component $xP^+$, where  \mbox{$ x = Q^2/2P \cdot q$} is the Bjorken variable, \mbox{$q^2 = -Q^2$} is the spacelike virtuality of the photon, and $P$ is the target momentum. To leading  order in the operator product expansion, the associated probability amplitude $q_f(x)$, characteristic of the target,  is  given by the  correlator~\cite{Jaffe:1985je,Jaffe:1983hp} 
\begin{equation}
q_f(x) = \frac{1}{4 \pi} \int  d z^- e^{i x P^+ z^- } 
\langle \pi(P) | \bar{\psi}_f (z^-) \gamma^+ \psi_f(0) | \pi(P) \rangle ~~~, 
\label{Mink_dis_x}
\end{equation}
where $f$ is a flavor label.
Here the 4-vector components that arise naturally are \mbox{$a^\pm = (a^0 \pm a^3)/\sqrt{2}$}.      This  probability amplitude is invariant, it  can easily be given a manifestly covariant formulation, and its interpretation is perhaps simplest in the infinite momentum frame where $q_f(x)$  is the quantum mechanical probability that a single parton  has momentum fraction 
$x$~\cite{Ellis:1991qj}.   Note that \mbox{$q_f(x) = - q_{\bar{f}}(-x) $}, and that the valence quark amplitude is \mbox{$ q^v_f(x) = q_f(x) - q_{\bar{f}}(x)$}.   It follows from Eq.~(\ref{Mink_dis_x}) that 
\mbox{$ \int_0^1 dx \,  q^v_f(x) = $} \mbox{$ \langle \pi(P) | J^+_f(0) | \pi(P) \rangle /2P^+ = F_\pi(0) = 1$}.   Approximate treatments should at least preserve  vector current conservation to automatically obtain the correct normalization for valence quark number.

In a momentum representation, $q_f(x)$ can be written as the evaluation of a special Feynman diagram~\cite{Jaffe:1985je,Mineo:1999eq}
%the + - notation in next eqn can have the 1/sqrt{2} or not
\begin{equation}
q_f(x) = \frac{1}{2} \int_k^\Lambda \delta(k^+ - xP^+) \; {\rm tr}_{\rm cd} [ \gamma^+ \, G(k,P) ]~~~,  
\label{Mink_dis_k}
\end{equation}
where ${\rm tr}_{\rm cd}$ denotes a color and Dirac trace, and $G(k,P)$ represents the forward 
$\bar q$-target scattering amplitude.   In ladder-rainbow truncation, which treats only the valence $q \bar q$ structure of the pion, Eq.~(\ref{Mink_dis_k}) yields  the explicit form
\begin{equation}
q^v_f(x) = \frac{i}{2} \int_p^\Lambda {\rm tr}_{\rm cd} [ \Gamma_\pi(p,P) \, S(p)\, \Gamma^+(p;x) \, S(p) \,\Gamma_\pi(p,P)\, S(p-P) ]~~~,  
\label{Mink_dis_LR_Ward}
\end{equation}
where $\Gamma^+(p;x)$ is a generalization of the dressed vertex that a zero momentum photon has with a quark.  It satisfies the usual  inhomogeneous BSE integral equation  (here with a LR kernel) except that the inhomogeneous term is 
$\gamma^+ \, \delta(p^+ - xP^+)$.   This selection of LR dynamics exactly parallels the symmetry-preserving dynamics of the corresponding treatment of the pion charge form factor at \mbox{$q^2 = 0 $} wherein the vector current is conserved by use of ladder dynamics at all three vertices and rainbow dynamics for all 3 quark propagators~\cite{Maris:2000sk,Maris:1999bh}.   Here the number of valence $u$-quarks (or $\bar d$) in the pion is automatically unity since the structure of Eq.~(\ref{Mink_dis_LR_Ward}), along with the canonical  normalization of the $q \bar q$ BS amplitude $\Gamma_\pi(p,P)$,  ensures \mbox{$ \int_0^1 dx \,  q^v_u(x) = 1$} because $ \int_0^1 dx \, \Gamma^+(p;x)$ gives the Ward Identity vertex.  

Eq.~(\ref{Mink_dis_LR_Ward}) is in Minkowski metric so as to satisfy the constraint on $p^+$, but LR dynamical information on the various non-perturbative elements such as $S(p)$ and  $\Gamma_\pi(p,P)$ is available only in Euclidean metric~\cite{Maris:1999nt}.   Since $q_f(x)$ is obtained from the hadron tensor $W^{\mu \nu}$ which in turn can be formulated from the discontinuity  \mbox{$T^{\mu\nu}(\epsilon) -  T^{\mu\nu}(-\epsilon)$}, we observe that all enclosed singularities from the difference of Wick rotations cancel except for the cut that defines the object of interest.    With use of numerical solutions for dressed propagators  and BS amplitudes, that give an accurate account of light quark hadrons, our DIS calculations  significantly extend  the exploratory study made in Ref.~\cite{Hecht:2000xa}.   That work employed phenomenological parameterizations of these elements.  

In Fig.~\ref{fig:pi_DSE+ratio} we display our DSE result for the valence $u$-quark distribution evolved to $Q^2 = (5.2~{\rm GeV})^2$ in comparison with $\pi N$ Drell-Yan data~\cite{Conway:1989fs} with a scale quoted as  $Q^2 > (4.05~{\rm GeV})^2$.   We also compare with  a recent NLO reanalysis of the data at scale $Q^2 = (5.2~{\rm GeV})^2$.   The distribution at the model scale $Q_0^2$ is evolved higher by leading order DGLAP.    The model scale is found to be \mbox{$Q_0 = 0.57 $}~GeV by matching the $x^n$ moments for $n=1,2,3$ to the experimental values given independently at (2~GeV)$^2$~\cite{Sutton:1991ay}.   
Our momentum sum rule result \mbox{$\int_0^1 dx x(u_\pi + \bar{d}_\pi ) = $} \mbox{$0.74 $} at $Q_0$ clearly show that in a covariant approach the retardation effects of one gluon exchange
assign some of the momentum to gluons.   The corresponding momentum sum for the kaon is $0.76$.

The ratio $u_K/u_\pi$ measures the dynamical effect of the local environment.   In the kaon, the 
$u$-quark is partnered with a significantly heavier partner than in the pion and this shifts the probability to relatively lower $x$ in the kaon.   Our preliminary DSE model 
calculation~\cite{Nguyen:inprep10,Nguyen_PhD09,Holt:2010vj} is shown in Fig.~\ref{fig:pi_DSE+ratio}  along with available Drell Yan  data~\cite{Badier:1980jq,Badier:1983mj}.    Here we include only the leading two invariants of the pion BS amplitude, $E(q, P)$ and $F(q, P)$, where $q$ is $q \bar q$ relative momentum.   For both amplitudes only the lowest Chebychev moment in $q\cdot P$ is employed .   This variable does not occur  in static quantum mechanics, nor in the Nambu--Jona-Lasinio point-coupling field theory model~\cite{Shigetani:1993dx} which also neglects the $q^2$ dependence.   We do not make such a point meson approximation here;  the $q^2$ dependence comes from the BSE solutions.   Nevertheless, the essential features of the ratio $u_K/u_\pi$  are adequately reproduced by a generalized Nambu--Jona-Lasinio model~\cite{Holt:2010vj}.

%%%%%%%%%%%%%%%%%%%%%%%%%%%%%%%%%%%%%%%%%%%%%%%%
%% BACKMATTER
%%%%%%%%%%%%%%%%%%%%%%%%%%%%%%%%%%%%%%%%%%%%%%%%

\begin{theacknowledgments}
The authors would like to thank  C. D. Roberts, S. J. Brodsky and I. Cloet for helpful conversations
and suggestions.   PCT thanks the Department of Physics, University of Sinaloa, for organizing a successful workshop, and providing warm hospitality.  This work has been partially supported by  the U.S. National Science Foundation under grant no. \ PHY-0903991, part of which constitutes  USA-Mexico collaboration funding in partnership with the Mexican agency CONACyT.
\end{theacknowledgments}



\bibliographystyle{aipproc}   % if natbib is available

%pct%\bibliography{sample}
\bibliography{refsPM,refsPCT,refsCDR,refs,refsMAP}

\IfFileExists{\jobname.bbl}{}
 {\typeout{}
  \typeout{******************************************}
  \typeout{** Please run "bibtex \jobname" to optain}
  \typeout{** the bibliography and then re-run LaTeX}
  \typeout{** twice to fix the references!}
  \typeout{******************************************}
  \typeout{}
 }

\end{document}